\documentclass[aps,twocolumn,superscriptaddress,prl]{revtex4-2}

\usepackage{bm,graphicx,textcomp,amssymb,amsmath,dcolumn,color,xcolor,hyperref}

\newcommand{\ket}[1]{\left|#1\right>}
\newcommand{\bra}[1]{\left<#1\right|}
\newcommand{\braket}[2]{\left<#1|#2\right>}
\newcommand{\nn}{\nonumber\\}

\newcommand{\f}[1]{\mbox{\boldmath$#1$}}

\newcommand{\bea}{\begin{eqnarray}}
\newcommand{\ea}{\end{eqnarray}}
\newcommand{\eea}{\end{eqnarray}}
\newcommand{\ord}{\,{\cal O}}

\begin{document}

\title{Faster than square-root run-time 
in adiabatic quantum algorithms for Exact Cover 3}

\author{Gernot Schaller}

\affiliation{Helmholtz-Zentrum Dresden-Rossendorf,
Bautzner Landstra{\ss}e 400, 01328 Dresden, Germany}

\author{Ralf Sch\"utzhold}

\affiliation{Helmholtz-Zentrum Dresden-Rossendorf,
Bautzner Landstra{\ss}e 400, 01328 Dresden, Germany}
\affiliation{Institut f\"ur Theoretische Physik,
Technische Universit\"at Dresden, 01062 Dresden, Germany}

\date{\today}

\begin{abstract}
By intertwining ideas from sequential and adiabatic quantum computing,
we propose a quantum algorithm for the problem class Exact Cover 3,
which is a sub-class of 3-satisfiability (3-SAT) problems and NP
(nondeterministic polynomial time) complete.
We find a quantum run-time which is significantly faster than the 
square-root run-time $N\to N^{1/2}$ of the Grover algorithm. 
Depending on the structure of the problem (e.g., connectivity of clauses), 
the run-time scaling may range from approximately third root $N\to N^{1/3}$ 
to nearly fourth root $N\to N^{1/4}$, possibly even faster.
\end{abstract}

\maketitle

\paragraph{Introduction}

After the pioneering works of Feynman~\cite{feynman1982a},
Shor~\cite{shor1997a}, Grover~\cite{grover1997a} and
many others (see, e.g.,~\cite{nielsen2000}), it has been a long standing question of
which advantages quantum algorithms can offer for solving classical problems.
The Grover search algorithm~\cite{grover1997a} for finding a specific item
(e.g., the zero of a complicated function) in an unsorted database
containing $N$ entries offers a quadratic speed-up $N\to N^{1/2}$
in comparison to a classical brute-force search.
In the absence of any structure in the problem, it can be shown~\cite{bennett1997a}
that this quadratic quantum speed-up $N\to N^{1/2}$ is optimal, i.e.,
some additional information about the problem or its solution is required
for achieving further speed-up.
The most prominent example is Shor's factoring algorithm~\cite{shor1997a}
with an exponential speed-up $N\to\ord([\ln N]^3)$.

Typically NP (nondeterministic polynomial time) problems are considered
promising candidates for quantum algorithms~\cite{choi2011a}.
Simply put, the solution of an NP problem is easy to verify but may be
hard to find~\cite{cormen1990}.
Examples are factoring, the eight-queens problem or the sudoku puzzles
in newspapers.
A sub-class of this complexity class are the NP complete problems, 
onto which all other NP problems can be mapped efficiently.
In this sense, NP complete problems are 
the hardest problems in NP.
Note that factoring, for example, is not NP complete.
In this Letter, we shall focus on the problem class Exact Cover~3 (EC3)
which is NP complete.

In the usual scheme of sequential quantum computing, the final state
$\ket{\psi_{\rm out}}$ containing the solution to the problem 
is obtained after applying a sequence of quantum gates to some initial
state $\ket{\psi_{\rm in}}$.
In contrast, adiabatic quantum computing is based on the idea of
constructing a suitable Hamiltonian $\hat H^{\rm out}$ such that
the sought-after state $\ket{\psi_{\rm out}}$ is its ground state,
see, e.g.,~\cite{albash2018a}.
To arrive at the desired state $\ket{\psi_{\rm out}}$,
the strategy 
is to start with an initial Hamiltonian $\hat H^{\rm in}$,
whose ground state $\ket{\psi_{\rm in}}$ 
is known and can be prepared easily, and then to
slowly transform from $\hat H^{\rm in}$ to $\hat H^{\rm out}$.
Usually, a direct 
interpolation scheme with a switching function $f(t)$ is considered 
\bea
\label{adiabatic-original}
\hat H(t)=[1-f(t)]\hat H^{\rm in}+f(t)\hat H^{\rm out}
\,,
\ea
but we employ a different scheme here,
see also~\cite{roland2003a,mao2005b,siu2007a,schaller2008b,
hastings2008a,rezakhani2009a,sun2013a,hofmann2014a,zeng2016a,matus2023a}.

Via the adiabatic theorem~\cite{born1928a}, the run-time required for
adiabatic evolution (i.e., staying close to the instantaneous ground
state all the time) is determined by the energy gap $\Delta E$
between the ground and the first excited state(s), which typically
becomes quite small at some point.
Estimating this minimum energy gap $\Delta E$ and thereby the required 
run-time for general problems is still an open question.
It has been suggested that adiabatic quantum algorithms might offer
an exponential speed-up for NP complete problems~\cite{farhi2001a}, but this
claim is debated~\cite{vandam2001a,znidaric2006a,young2008a,altshuler2010a,
young2010a,hen2011a}.
For example, based on Anderson like localization phenomena, it has
been argued~\cite{altshuler2010a} that such adiabatic quantum algorithms
should not be
faster than the square-root Grover run-time $N\to N^{1/2}$ for large $N$, 
at least for the conventional scheme~\eqref{adiabatic-original}.
For a specific class of Hamiltonians $\hat H^{\rm in}$ for which the
minimum gap $\Delta E_{\rm min}$ could be calculated or estimated
analytically, a square-root run-time 
has been found~\cite{roland2002a,znidaric2006a}.

In the following, we intertwine ideas from sequential and adiabatic
quantum computing and propose adiabatic quantum algorithms deviating
from the conventional scheme~\eqref{adiabatic-original}.
We show that the achievable run-time is much faster than in the Grover case 
$N\to N^{1/2}$, such as $N\to N^{1/3}$ or $N\to N^{1/4}$.

\paragraph{Exact Cover 3}

The problem class~\cite{karp1972a,farhi2001a} Exact Cover~3 (EC3) 
can be formulated in the following way:
For $n$ bits $z_\mu\in\{0,1\}$, we seek a sequence $w$
(referred to as the solution) such as $w=10\dots01$ 
(where we would have $z_1^w = 1, z_2^w=0, \ldots, z_{n-1}^w=0, z_n^w=1$)
which satisfies $m$
constraints $z_\mu^w+z_\nu^w+z_\rho^w=1$ (referred to as clauses).
The arrangement of these $m$ triples $\mu,\nu,\rho$ then defines
the specific problem to be solved~\footnote{
Note that two-bit clauses of the form $z_\mu+z_\nu=1$ impose a direct
relation between these two bits $z_\mu=1-z_\nu=\bar z_\nu$ and thus
reduce the complexity of the problem.
To account for this option, one could also include clauses of the form
$\bar z_\mu+z_\nu+z_\rho=1$.}.

Here we assume that the problem has a unique solution $w$
(or at least a very small number of solutions) since these
are typically the hard problems~\cite{farhi2001a,kalapala2008a}.
Thus, we need at least $m\ge n/3$ clauses to cover all bits.
Excluding the case of contradicting clauses (such that no solution exists), 
there would 
still be a huge number of possible solutions for this small $m$.
Adding more and more clauses then restricts the number of solutions
further and further, until this number drops down to unity
(or a small number) at a percolation like threshold at a ratio of $m/n$
between 60~\% and 80~\%~\cite{kalapala2008a,young2008a,moore2016a,zhang2022b}.
To be specific, we take $m\approx 2n/3$ in the following~\cite{kalapala2008a}.

As a result, each bit is involved in two clauses on average.
Actually, this is also the most relevant case: 
Obviously, a bit which is not involved in any clause can be omitted,
such that every bit must be involved in at least one clause.
If a bit is involved in only one clause, it could 
also be omitted and the remaining two bits then would 
have to satisfy the two-bit clause $z_\mu z_\nu=0$.
In the other direction, a bit involved in many clauses does also
allow us to reduce the complexity by a distinction of cases:
If it is 1, all other bits involved in these clauses must be 0;
if it is 0, these three-bit clauses turn to two-bit clauses
$z_\mu+z_\nu=1$ which also reduce the complexity. 
Thus, in the following, we assume that almost all bits are involved
in two clauses, with some rare exceptions (e.g., one or three clauses),
see also~\cite{schaller2010a}.

An even more compelling argument applies to pairs of bits.
If two bits, say $z_\mu$ and $z_\nu$, are involved in two clauses,
such as $z_\mu+z_\nu+z_\rho=1$ and $z_\mu+z_\nu+z_\lambda=1$, then the
remaining third bits must be equal $z_\rho=z_\lambda$,
which would again allow us to reduce the complexity.
Thus, we assume (without exceptions) that each pair shares at most 
one clause~\cite{hofmann2014a}.

\paragraph{Adiabatic Quantum Algorithm}

In order to construct a Hamiltonian whose ground state encodes
the sought after solution $w$ for Exact Cover 3, one option is~\cite{banuls2006a,schuetzhold2006b}
\bea
\label{Hamiltonian-ec3}
\hat H^{\rm EC3}
&=&
\frac{1}{4}
\sum\limits_{\mu\nu\rho}^{{\rm clauses}}
\left(\sigma^z_\mu + \sigma^z_\nu + \sigma^z_\rho - \f{1}\right)^2
\nn
&=&
m \f{1} - \sum\limits_\mu \frac{N_\mu}{2}\, \sigma^z_\mu + 
\sum\limits_{\mu\nu} \frac{T_{\mu\nu}}{4}\, \sigma^z_\mu \sigma^z_\nu
\,,
\ea
where the 
Pauli spin matrix 
$\sigma^z_\mu=(\ket{0}\bra{0}-\ket{1}\bra{1})_\mu$ 
acts on the qubit at position $\mu$. 
Here, $T_{\mu\nu}$ and $N_\mu$ denote the
number of clauses involving the pair $\mu$, $\nu$ of bits or the
single bit $\mu$, respectively.
As explained above, $T_{\mu\nu}$ is either zero or one while $N_\mu$
is typically two.
As a result, the Hamiltonian~\eqref{Hamiltonian-ec3} corresponds to a
frustrated anti-ferromagnet with local fields in a complex 
geometry~\footnote{
Other clauses such as $\bar z_\mu+z_\nu+z_\rho=1$ or $z_\mu z_\nu=0$
could be implemented in analogy via
$(\f{1}-\sigma^z_\mu+\sigma^z_\nu+\sigma^z_\rho)^2$
or $(\f{1}+\sigma^z_\mu)(\f{1}+\sigma^z_\nu)$, respectively.
They would change the explicit values of $T_{\mu\nu}$ and $N_\mu$,
but otherwise the structure remains the same.}.
The Hamiltonian~\eqref{Hamiltonian-ec3} is non-negative and the
solution $w$ to all clauses is its ground state, i.e.,
the kernel $\hat H^{\rm EC3}\ket{w}=0$.

Following the usual scheme of adiabatic quantum computing,
one could start from a suitable initial Hamiltonian such as
$\hat H^{\rm in}=\pm\sum_\mu\sigma^x_\mu$
and then interpolate from
$\hat H^{\rm in}$ to $\hat H^{\rm out}=\hat H^{\rm EC3}$
via~\eqref{adiabatic-original}, see, e.g.,~\cite{farhi2001a,roland2002a}.

\begin{figure}
\includegraphics[width=0.4\textwidth]{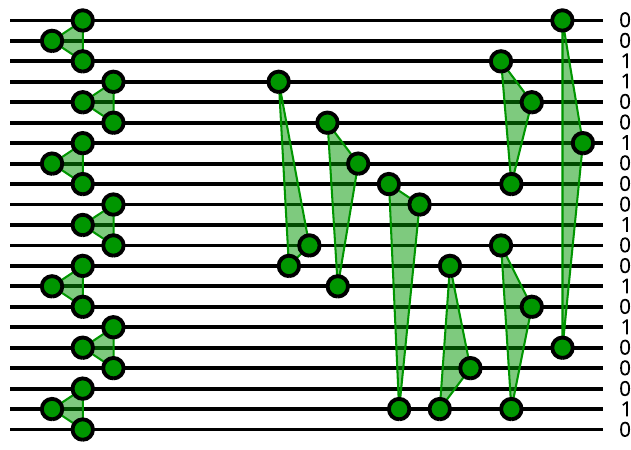}
\includegraphics[width=0.45\textwidth]{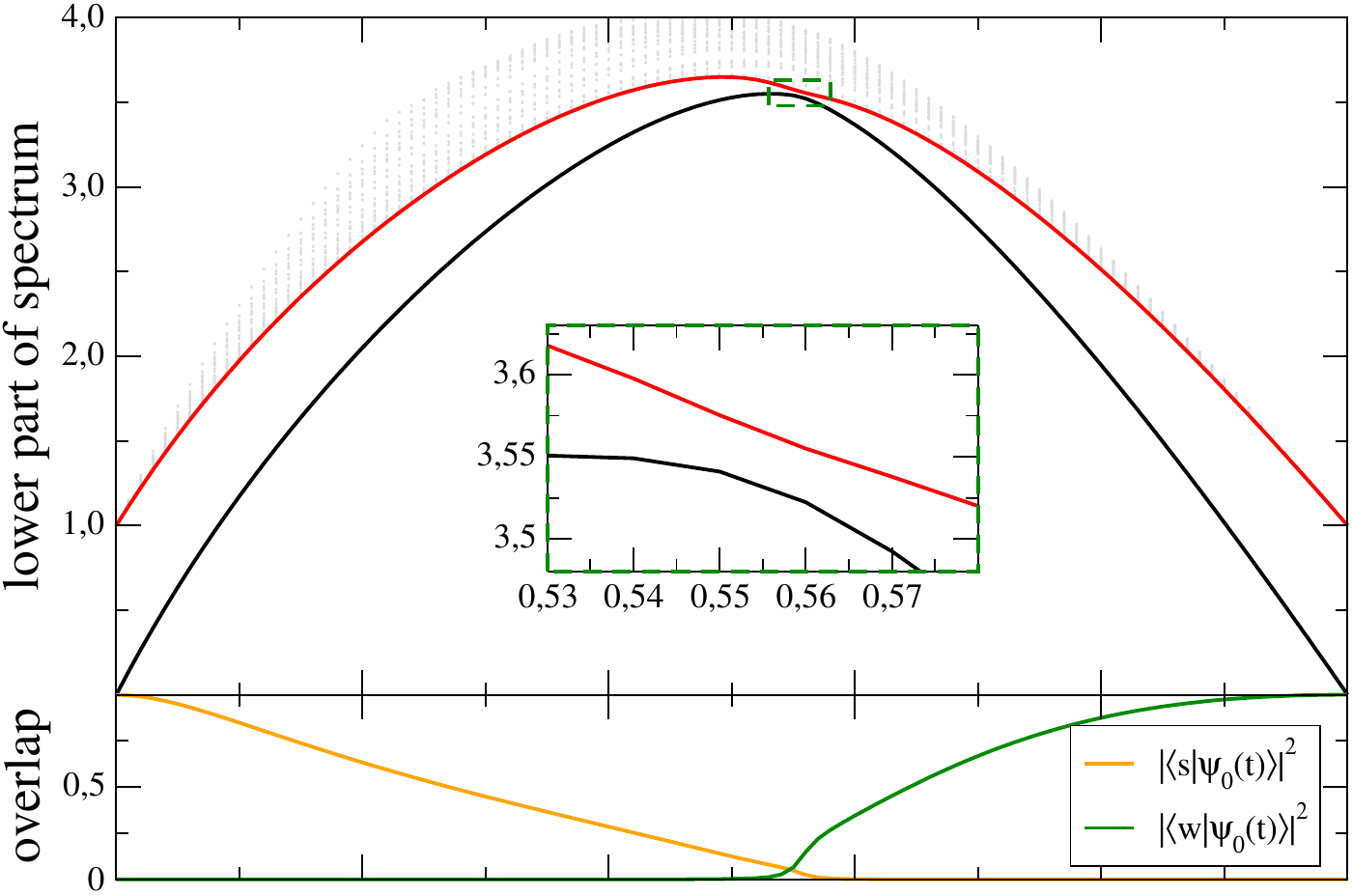}
\includegraphics[width=0.45\textwidth]{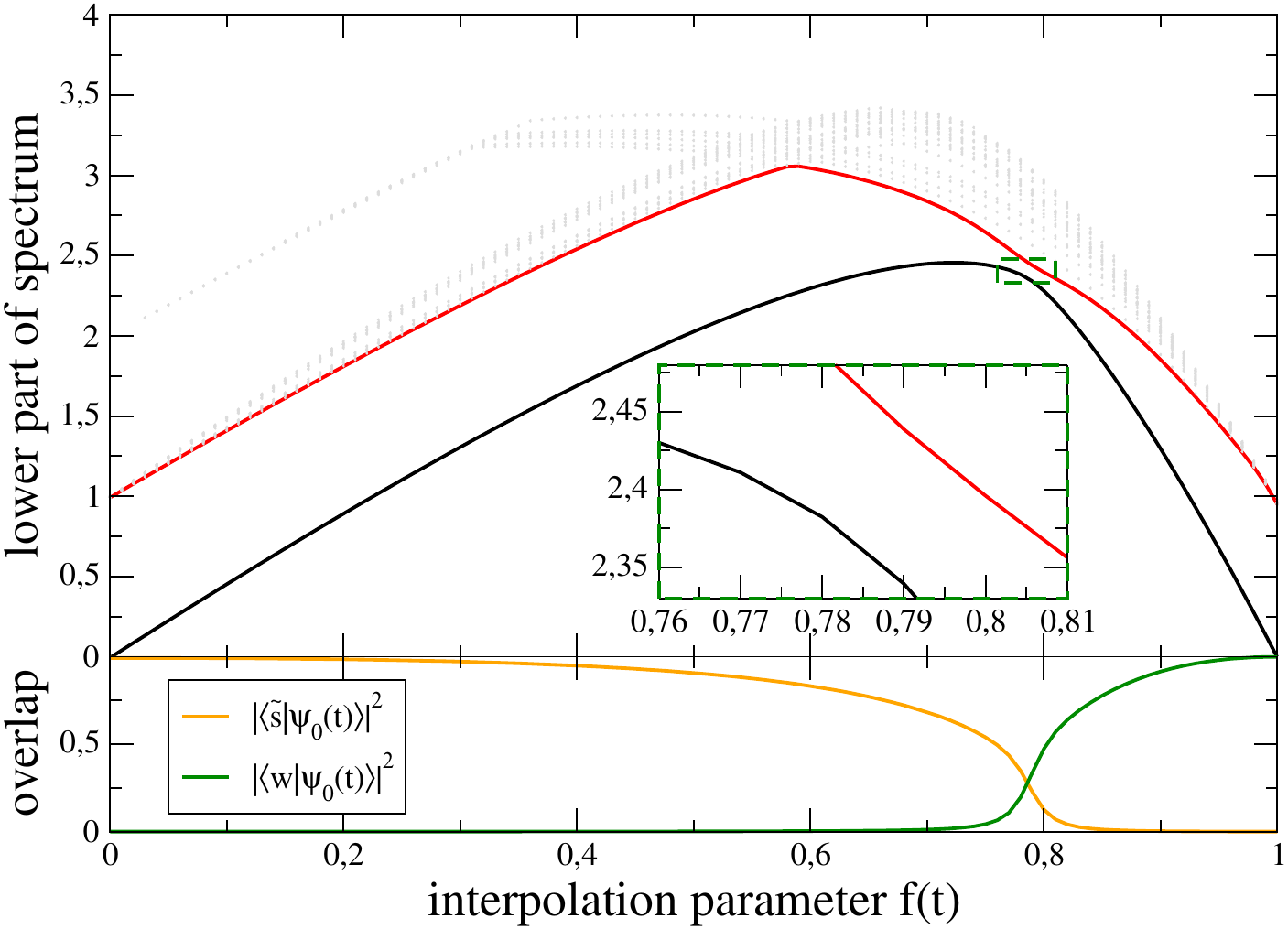}
\caption{\label{FIG:intpolAB}
Comparison between the conventional scheme~\eqref{adiabatic-original}
and the step-wise algorithm~\eqref{interpolation} proposed here for
a specific example.
The top panel depicts the considered Exact Cover~3 instance with
21~bits (black horizontal lines) and 14~clauses (green triangles).
The $m_1=7$ non-overlapping clauses are arranged on the left-hand
side and the remaining $m_2=7$ clauses to their right.
The unique solution $w$ is given in the column on the right-hand side.
The middle panel shows the lowest two eigenvalues
(black and red curve) of the Hamiltonian~\eqref{adiabatic-original}
as a function of the interpolation parameter $f(t)$ as well as the
probability overlap between the instantaneous ground state
$\ket{\psi_0(t)}$ and the initial ground state
$\ket{s}=\ket{+\ldots+}$ (yellow curve) and the final ground state
$\ket{w}$ (green curve), respectively.
In analogy, the bottom panel displays these values for the
step-wise algorithm~\eqref{interpolation}, where the initial
ground state $\ket{\tilde s}$ 
is the reduced superposition state~\eqref{s-state}.
For the latter, the minimum gap is around a factor of three larger
and the Landau-Zener two-mode description~\eqref{Landau-Zener-general}
a much better approximation.}
\end{figure}

However, here we propose a different scheme.
Instead of switching on all clauses simultaneously, we first
split up the set of all $m$ clauses into two sub-sets of $m_1$
and $m_2$ clauses, respectively, such that $m=m_1+m_2$ and the
first $m_1$ clauses are non-overlapping.
Here non-overlapping means that none of the first $m_1$ clauses
share the same qubit, i.e., no qubit involved in one of the first
$m_1$ clauses is involved in any other of these $m_1$ clauses
(but typically in one of the other $m_2$ clauses).
Thus, these first $m_1$ clauses are disconnected and we may
switch them on (or implement them) independently of each other, 
see Fig.~\ref{FIG:intpolAB}.
More precisely, we split up the Hamiltonian~\eqref{Hamiltonian-ec3}
into these two sub-sets
$\hat H^{\rm EC3}=\hat H^{\rm EC3}_1+\hat H^{\rm EC3}_2$
and first switch on $\hat H^{\rm EC3}_1$.
The ground-state manifold of $\hat H^{\rm EC3}_1$, i.e.,
its kernel, contains all computational basis states $\ket{z}$
which satisfy the first $m_1$ clauses, i.e.,
$\hat H^{\rm EC3}_1\ket{z}=0$.
Preparing an initial state $\ket{\psi_{\rm in}}$ within this manifold
(as specified below) then amounts to implementing the first $m_1$ clauses.

\paragraph{Initial State}

Within this manifold, the initial state $\ket{\psi_{\rm in}}$ 
is chosen such that it coincides with a W state
\bea
\label{W-state}
\ket{{\rm W}}=\frac{\ket{001}+\ket{010}+\ket{100}}{\sqrt{3}}
\ea
for each triple of qubits involved in one of the first $m_1$ clauses,
i.e., a superposition of the three solutions of that clause with
equal weight~\footnote{
For a clause $\bar z_\mu+z_\nu+z_\rho=1$ instead of
$z_\mu+z_\nu+z_\rho=1$, we would just flip the first qubit $\ket{z_\mu}$,
i.e., the state~\eqref{W-state} would become
$(\ket{101}+\ket{110}+\ket{000})/\sqrt{3}$.
For a two-bit clause $z_\mu z_\nu=0$, the corresponding superposition
state reads $(\ket{10}+\ket{01}+\ket{00})/\sqrt{3}$.}. 
For qubits which are not involved in any of the first $m_1$
clauses, we take the plus state $\ket{+}=(\ket{0}+\ket{1})/\sqrt{2}$.
The desired initial state after implementing the first $m_1$ clauses 
would then be 
the tensor product of the W states~\eqref{W-state} for the triples
involved in the first $m_1$ clauses 
and the plus states $\ket{+}$ for the remaining single qubits.
As a result, it is the reduced superposition state
\bea
\label{s-state}
\ket{\psi_{\rm in}}
=
\ket{\tilde s}
=\frac{1}{\sqrt{N_1}}\sum\limits_{z\in\{0,1\}^{\otimes n}}^{{m_1\rm \;clauses}}
\ket{z}
\ea
of all the $N_1$ solutions of the first $m_1$ clauses, i.e., those
computational basis states obeying $\hat H^{\rm EC3}_1\ket{z}=0$.
Note that, for a sufficiently large number $m_1$ of first clauses,
this number $N_1$ is much smaller $N_1\ll N$ than the total number
$N=2^n$ of computational basis states.

In order to stabilize the initial state~\eqref{s-state} as the unique 
ground state of the initial Hamiltonian $\hat H^{\rm in}$,
we may add $\hat H^{\rm in}=\hat H^{\rm EC3}_1+\hat H^{\rm mix}$
a suitable scrambling or mixing Hamiltonian $\hat H^{\rm mix}$
of the form
\bea
\label{scrambling}
\hat H^{\rm mix}=
\sum\limits_{\mu\nu}^{m_1 \rm \;clauses}
\frac{T_{\mu\nu}}{2} 
\left(\f{1}-\sigma^x_\mu\sigma^x_\nu-\sigma^y_\mu\sigma^y_\nu-\sigma^z_\mu \sigma^z_\nu\right)
\,,
\ea
where the sum over $\mu$ and $\nu$ is restricted to those pairs of qubits
which are connected by one of the first $m_1$ clauses.
Note that the above form~\eqref{scrambling} has already been suggested in~\cite{schuetzhold2006b} and
encodes SWAP operations on the qubit pairs $\mu$ and $\nu$.
For the remaining single qubits $\ket{z_\rho}$ not involved in any of the
first $m_1$ clauses, we may add the usual term 
$(\f{1}-\sigma_\rho^x)/2$ 
in order to stabilize their plus states.
%

\paragraph{Adiabatic Evolution}

In order to evolve from the initial state~\eqref{s-state} to the sought-after
solution state $\ket{w}$ which satisfies all clauses, we have to switch on
the remaining $m_2$ clauses.
The idea is to stay in the ground-state manifold of $\hat H^{\rm EC3}_1$
(i.e., its kernel) such that we may keep $\hat H^{\rm EC3}_1$ switched on.
Then, to eliminate all states in this ground-state manifold which violate
any of the remaining $m_2$ clauses, we slowly switch on $\hat H^{\rm EC3}_2$
and reduce the impact of $\hat H^{\rm mix}$ accordingly, e.g., with $f(t)\in[0,1]$
\bea
\label{interpolation}
\hat H(t)=\hat H^{\rm EC3}_1+[1-f(t)]\hat H^{\rm mix}+f(t)\hat H^{\rm EC3}_2
\,.
\ea
According to the adiabatic theorem, the corresponding quantum evolution
stays close to the (instantaneous) ground state provided that it is slow
enough, where the allowed interpolation speed $df/dt$ is set by the
minimum gap $\Delta E$ between the (instantaneous) ground and first
excited state(s).

To a much stronger extent than in the traditional scheme 
(see also~\cite{farhi2001a} and Fig.~\ref{FIG:intpolAB}),
we have an avoided level crossing
between the (instantaneous) ground and first excited state at some point
of the evolution, where $\Delta E$ becomes very small.
To estimate the minimum gap $\Delta E$, we focus on this avoided level
crossing and neglect all higher lying states, such that we may approximate
it as a Landau-Zener problem.
Since the major change of the ground state occurs in the vicinity of this
avoided level crossing, we neglect the quantum evolution before and after,
such that we may focus to the effective Landau-Zener sub-space spanned by
the initial and final ground states $\ket{\tilde s}$ and $\ket{w}$ or,
more precisely $\ket{\tilde s_\perp}\propto
\ket{\tilde s}-\ket{w}\braket{w}{\tilde s}$
and $\ket{w}$.
In this Landau-Zener sub-space, the effective Hamiltonian can be represented 
as
\bea
\label{Landau-Zener-general}
\hat H\to
\left(
\begin{array}{cc}
\bra{w}\hat H\ket{w} & \bra{w}\hat H\ket{\tilde s_\perp}
 \\
\bra{\tilde s_\perp}\hat H\ket{w} &
\bra{\tilde s_\perp}\hat H\ket{\tilde s_\perp}
\end{array}
\right)
\,,
\ea
where the avoided level crossing occurs when the diagonal terms coincide
$\bra{w}\hat H\ket{w}=\bra{\tilde s_\perp}\hat H\ket{\tilde s_\perp}$.
Since the solution $\ket{w}$ is annihilated by both $\hat H^{\rm EC3}_1$
and $\hat H^{\rm EC3}_2$ while changing any of the qubits with
$\hat H^{\rm mix}$ is typically not a solution anymore, we have
$\bra{w}\hat H\ket{w}=0$ for all $f(t)$.
Similarly, $\ket{\tilde s_\perp}$ is annihilated by $\hat H^{\rm EC3}_1$
such that the value of $f(t)$ at the avoided level crossing is determined
by a balance of
$\bra{\tilde s_\perp}\hat H^{\rm mix}\ket{\tilde s_\perp}$
and
$\bra{\tilde s_\perp}\hat H^{\rm EC3}_2\ket{\tilde s_\perp}$
which depends on $m_1$ and $m_2$ etc.


The minimum energy gap $\Delta E_{\rm min}$ at the avoided level
crossing is then set by the off-diagonal terms
$\bra{w}\hat H\ket{\tilde s_\perp}$.
Again, as $\ket{w}$ is annihilated by both $\hat H^{\rm EC3}_1$
and $\hat H^{\rm EC3}_2$, the off-diagonal terms are generated by the
scrambling Hamiltonian~\eqref{scrambling} via
$\bra{w}\hat H^{\rm mix}\ket{\tilde s_\perp}$.
Quite intuitively, such a mixing is required for the transition from a
wrong solution contained in $\ket{\tilde s_\perp}$ to the correct solution
$\ket{w}$.
When acting on the solution $\ket{w}$, the terms in $\hat H^{\rm mix}$
may swap a pair of qubits involved in any of the first $m_1$ clauses or
flip one of the remaining qubits.
Therefore, unless the swapping operation acts on two bits in the
solution $w$ which both happen to be zero
(they cannot both be one since they are involved in a clause),
the resulting state will be different from $\ket{w}$, but it will
be contained in the reduced superposition state $\ket{\tilde s_\perp}$.
Consequently, the off-diagonal terms scale (at least) as
$\bra{w}\hat H\ket{\tilde s_\perp}=\ord(1/\sqrt{N_1})$
and thus the minimum energy gap $\Delta E_{\rm min}$ also scales as
$\Delta E_{\rm min}=\ord(1/\sqrt{N_1})$, see Fig.~\ref{FIG:intpolAB}.
This, in turn, sets the required adiabatic run-time which scales
as $\ord(\sqrt{N_1})$ and thus may be much faster than the
Grover run-time $\ord(\sqrt{N})$ yielding a quadratic speed-up.
An alternative justification for the $\ord(\sqrt{N_1})$ scaling is 
provided in the Appendix. 

\paragraph{Quantum run-time scaling}

The achievable quantum run-time 
depends on $N_1$, i.e., on the number
$m_1$ of clauses to be switched on in the first step.
For a lower bound on this number, let us consider the following scheme:
First, we pick a clause at random and assign it to the set
of these $m_1$ clauses.
Then we take those clauses (typically three) which share one  
qubit with this first clause, i.e., overlap with it, and assign them to
the set of $m_2$ clauses to be switched on in the second step.
Since the remaining set of $m-4$ clauses is non-overlapping with
the first one, we can repeat the same sequence.
Actually, as in subsequent steps some clauses may already have been sorted 
to the $m_2$ set, the ratio of $m_2$ to $m_1$ clauses may be smaller
than three, such that we eventually
have at least $m_1 \ge m/4$ 
non-overlapping clauses for initial state preparation 
and less than $m_2 \le 3m/4$  
for the adiabatic algorithm~\eqref{interpolation}. 
For each of these $m_1$ clauses, the number of allowed solutions
is reduced by a factor of 3/8, i.e., the three involved qubits are
effectively reduced to ``one and a half'' qubits.
Thus $N_1=(3/8)^{m_1}N\lesssim(3/8)^{m/4}N$ 
and inserting $m\approx2n/3$ (as discussed above), we find a reduction of
$N_1\approx(3/8)^{n/6}N\approx(0.85)^nN=(1.7)^n$ instead of
$N=2^n$.
Since the run-time scales with $\sqrt{N_1}$, this amounts to
an almost third root run-time $N\to N^{1/3}$.

An even faster run-time can be reached if $m_1$ is larger.
Depending on the connectivity of the graph describing all bits
and clauses, one could find more non-overlapping clauses,
up to the
maximum number $m_1 \le m_1^{\rm max}=n/3$ 
(that one would obtain when the $m_1$ clauses cover all qubits).
For this upper bound, an analogous calculation yields an
approximately fourth root run-time $N\to N^{1/4}$.

\paragraph{Clusters of clauses}

Even shorter run-times can be reached if we allow for finite clusters of 
clauses in the set of $m_1$ clauses to be switched on first. 
As the simplest example, let us consider two clauses which share one qubit. 
Then the W state~\eqref{W-state} has to be adapted accordingly to a 
superposition state of the five allowed solutions for the five involved 
qubits. 
As a result, the factor of $3/8$ for a single clause is now reduced to 
$5/32$ for these two connected clauses. 
In analogy to the scheme described above, we may place this pair of clauses 
into the first set $m_1$ and the clauses (typically four) overlapping with 
this pair into the second set $m_2$. 
In this way, the ratio between clauses put in $m_1$ and $m_2$, respectively, 
is also improved from $1/3$ to $1/2$. 
Thus, the effective reduction factor per qubit of $0.85$ found above is 
now improved to $0.66$. 
How far one can push this scheme (e.g., with larger clusters) 
again depends on the 
connectivity of the graph describing all bits and clauses.

\paragraph{Conclusions}

We propose an adiabatic quantum algorithm for the NP complete problem class Exact 
Cover 3. 
In contrast to the usual scheme, we do not switch on all clauses 
simultaneously, but proceed in two steps. 
In the first step, a sub-set of $m_1$ clauses is implemented, while the 
remaining $m_2$ clauses are switched on adiabatically in the second step. 
In this way, we achieve a run-time which is significantly faster than the 
standard square-root run-time $N\to N^{1/2}$ of the Grover algorithm, 
such as $N\to N^{1/3}$ or $N\to N^{1/4}$, possibly even faster. 

Our analytical estimate of the minimum energy gap and thus run-time 
evades major shortcomings of numerical approaches. 
On the one hand, they are limited to finite system sizes, which makes 
it hard to obtain the correct scaling in the large-$N$ limit. 
On the other hand, they cannot consider all possible instances for 
large $N$ and the study of random instances may also not include the 
hard cases which are required for NP completeness. 

Apart from the run-time, the space (e.g., number of required bits or 
qubits at a given time) of an algorithm is also an important criterion. 
For the adiabatic quantum algorithm proposed here, the space scales 
linearly in $n$. 
An example for a classical algorithm with such a linear scaling is 
the brute-force search, but this requires a very long run-time $\ord(N)$. 
One could also restrict the search space classically, but this would 
still require a run-time of order $N_1$ and is thus inferior to the 
quantum algorithm proposed here.
Classical algorithms with shorter run-times $\ord(N^\xi)$ with $\xi<1$
such as the ``dancing links''~\cite{knuth2000a,hoi2020a,peng2025a}
are recursive and require far
more space than $\ord(n)$. 

As a related point, the energy consumption of the computation is also 
an increasingly important criterion. 
If we assign a certain energy cost per gate operation to a classical 
algorithm, the total energy consumption would scale exponentially 
$\ord(N^\xi)$ with the number $n$ of bits or qubits. 
Adiabatic quantum algorithms could facilitate a much better energy 
efficiency since they can be operated by a slowly varying classical 
(e.g., magnetic) field such that the total energy cost could scale 
linearly $\ord(n)$. 

It should also be stressed here that we did not incorporate the 
effects of an environment such as losses and decoherence. 
Actually, this issue points to another potential advantage of 
the adiabatic scheme: 
Already without error correction, adiabatic quantum algorithms are
more robust against the impact of an environment than sequential
quantum algorithms~\cite{childs2001a}.
Note, however, this does not mean that one can omit the coupling 
to the environment completely, see, e.g.,~\cite{ahmadiniaz2026a}. 
In this regard, increasing the energy gap $\Delta E$ by the two-step 
procedure proposed here should help to make the quantum algorithm
more robust against decoherence. 

\acknowledgments

\paragraph{Acknowledgments}

Funded by the Deutsche Forschungsgemeinschaft
(DFG, German Research Foundation)
through the Collaborative Research Center SFB~1242
(Project-ID 278162697).
R.~S.~has benefited from the activities of COST Action CA23115:
Relativistic Quantum Information, funded  by COST
(European Cooperation in Science and Technology).




\bibliographystyle{unsrt} 
\bibliography{references}


\appendix 

\section{Appendix: Sequential Quantum Algorithm}

While suggestive, the arguments in the main text do not constitute a rigorous proof for
a faster than square-root run-time. 
In order to demonstrate that such a reduced run-time 
is indeed possible for this problem class, let us switch back to the scheme 
of sequential quantum computing.
To this end, we present a generalization of the Grover search algorithm.

As the first step, we construct the unitary transformation
$\hat U_w=\f{1}-2\ket{w}\bra{w}$ generating a reflection at the
plane perpendicular to the solution state $\ket{w}$.
One way of doing that is to introduce $m$ ancilla qubits $\ket{y_i}$,
one for each clause $i$ involving the three qubits
$z_\mu$, $z_\nu$ and $z_\rho$.
Starting in the state $\ket{y_i}=\ket{0}$, we apply the unitary
\bea
\label{ancilla}
\ket{z_\mu}\ket{z_\nu}\ket{z_\rho}\ket{y_i}\to
\ket{z_\mu}\ket{z_\nu}\ket{z_\rho}\ket{y_i\oplus c_i(z_\mu,z_\nu,z_\rho)}
\,,
\ea
where $c_i(z_\mu,z_\nu,z_\rho)=1$ if the clause $i$ is satisfied and zero otherwise.
After testing all $m$ clauses, we apply a phase flip provided that all ancilla
qubits are $\ket{1}$, e.g., with an additional phase ancilla $\ket{x}$ and the
unitary
\bea
\label{phase}
\ket{y_1}\dots\ket{y_m}\ket{x}\to
\ket{y_1}\dots\ket{y_m}\ket{x\oplus y_1y_2\dots y_m}
\,.
\ea
With the initial state $(\ket{0}-\ket{1})/\sqrt{2}$ for this phase ancilla qubit, this unitary~\eqref{phase} introduces a conditional phase flip. 
%
Finally, we repeat the sequence~\eqref{ancilla} for all $m$ clauses in order 
to set all ancilla qubits $\ket{y_i}$ back to $\ket{0}$.
As a result, we obtain a phase flip for the solution state $\ket{w}\to-\ket{w}$,
while all other states are unaffected.
This constitutes the first half $\hat U_w=\f{1}-2\ket{w}\bra{w}$ of the Grover
iteration (often referred to as the oracle).

The second half of Grover iteration
$\hat U_{\tilde s}=\f{1}-2\ket{\tilde s}\bra{\tilde s}$
is also a reflection but now on the plane perpendicular to the reduced
superposition state~\eqref{s-state}.
This can be implemented with fewer ancilla qubits, one for each of the
first $m_1$ clauses in Eq.~\eqref{s-state}.
For example, we may apply unitary transformations $\hat U_i$ to the three qubits
$\ket{z_\mu}$, $\ket{z_\nu}$ and $\ket{z_\rho}$ involved in the clause $i$
which map the W states~\eqref{W-state} to $\ket{111}$.
Then we may entangle these three qubits to the corresponding ancilla qubit
$\ket{y_i}$ in analogy to Eq.~\eqref{ancilla}
\bea
\label{ancilla-s}
\ket{z_\mu}\ket{z_\nu}\ket{z_\rho}\ket{y_i}\to
\ket{z_\mu}\ket{z_\nu}\ket{z_\rho}\ket{y_i\oplus z_\mu z_\nu z_\rho}
\,.
\ea
As a result, the ancilla qubit $\ket{y_i}$ is switched to $\ket{1}$
if these three qubits are in the state $\ket{111}$, i.e., if they were
in the W state~\eqref{W-state} originally.
Then we may apply the phase flip in analogy to Eq.~\eqref{phase},
but this time the $m$ ancilla qubits $\ket{y_1}\dots\ket{y_m}$ for all
clauses are replaced by the $m_1$ ancilla qubits for the first
$m_1$ clauses and the remaining qubits in the plus states
(e.g., after applying Hadamard gates).
As before, the final step consists in reversing the applied unitary
transformations $\hat U_i$ and the Hadamard gates as well
as~\eqref{ancilla-s} such that the qubits
$\ket{z_\mu}$ and the ancilla qubits $\ket{y_i}$ are set back to their
original states -- apart from the phase flip of the state $\ket{\tilde s}$,
while again all other states remain unaffected.
In this way, we may implement the second unitary
$\hat U_{\tilde s}=\f{1}-2\ket{\tilde s}\bra{\tilde s}$
which completes the Grover iteration.

As in the usual Grover algorithm, the consecutive application of the
two unitary operations $\hat U_{\tilde s}\hat U_w$ effectively
amounts to a rotation within the sub-space spanned by $\ket{w}$
and $\ket{\tilde s}$.
However, due to $\braket{w}{\tilde s}=1/\sqrt{N_1}$ instead of
$1/\sqrt{N}$, the rotation angle for each iteration is much larger
than in the usual Grover algorithm.
Thus, starting in the $\ket{\tilde s}$ state~\eqref{s-state}, we need
$\ord(\sqrt{N_1})$ iterations for reaching the $\ket{w}$ state,
i.e., the solution to our problem.

Note that the Grover algorithm within a reduced search space has
already been discussed in the literature~\cite{mandra2016a,ambainis2019a} though in a different context.
Furthermore, instead of the full phase flips in 
$\hat U_w=\f{1}-2\ket{w}\bra{w}$ and 
$\hat U_{\tilde s}=\f{1}-2\ket{\tilde s}\bra{\tilde s}$, 
one could also generate gradual phase rotations such as 
$\hat U^\varphi_w=\f{1}-\ket{w}\bra{w}+e^{i\varphi}\ket{w}\bra{w}$ 
and analogously for $\hat U_{\tilde s}^\varphi$ with some small 
phase $\varphi\ll1$. 
In this way, the repeated consecutive application of the
two unitary operations $\hat U_{\tilde s}^{\varphi_1}\hat U_w^{\varphi_2}$
with suitably chosen phases $\varphi_1$ and $\varphi_2$ yields the 
Trotter representation~\cite{suzuki1990a} of the effective 
Hamiltonian~\eqref{Landau-Zener-general} which brings us back to the 
adiabatic scheme. 
Since the minimum energy gap is quite small, the phases $\varphi_1$ 
and $\varphi_2$ corresponding to the Trotter discretization of time 
need not be very small.

\end{document}